\documentclass[prd,aps,reprint,superscriptaddress,floatfix,flushbottom,nobibnotes,nofootinbib,showpacs,amsmath]{revtex4-1}
\usepackage{amssymb}
\usepackage{graphicx}
\begin{document}
\title{
$\boldsymbol{\langle}\boldsymbol{x}\boldsymbol{\rangle}_{\boldsymbol{u}\boldsymbol{-}\boldsymbol{d}}$ from lattice QCD at nearly physical quark masses}
\author{Gunnar S.~Bali}\email{Corresponding author. E-mail: gunnar.bali@ur.de}
\affiliation{Institut f\"ur Theoretische Physik, Universit\"at Regensburg,
              93040 Regensburg, Germany}
\author{Sara~Collins}
\affiliation{Institut f\"ur Theoretische Physik, Universit\"at Regensburg,
              93040 Regensburg, Germany}
\author{Mridupawan~Deka}
\affiliation{Institut f\"ur Theoretische Physik, Universit\"at Regensburg,
              93040 Regensburg, Germany}
\author{Benjamin~Gl\"a\ss{}le}        
\affiliation{Institut f\"ur Theoretische Physik, Universit\"at Regensburg,
              93040 Regensburg, Germany}
\author{Meinulf~G\"{o}ckeler}        
\affiliation{Institut f\"ur Theoretische Physik, Universit\"at Regensburg,
              93040 Regensburg, Germany}
\author{Johannes Najjar}        
\affiliation{Institut f\"ur Theoretische Physik, Universit\"at Regensburg,
              93040 Regensburg, Germany}
\author{Andrea~Nobile}  
\affiliation{JSC, Research Centre J\"{u}lich, 52425 J\"{u}lich, Germany}
\author{Dirk~Pleiter}  
\affiliation{JSC, Research Centre J\"{u}lich, 52425 J\"{u}lich, Germany}
\affiliation{Institut f\"ur Theoretische Physik, Universit\"at Regensburg,
              93040 Regensburg, Germany} 
\author{Andreas Sch\"{a}fer}  
\affiliation{Institut f\"ur Theoretische Physik, Universit\"at Regensburg,
              93040 Regensburg, Germany}  
\author{Andr\'e~Sternbeck}  
\affiliation{Institut f\"ur Theoretische Physik, Universit\"at Regensburg,
              93040 Regensburg, Germany}  
\date{\today}
\begin{abstract}
We determine the second Mellin moment of the isovector quark
parton distribution function $\langle x\rangle_{u-d}$ from
lattice QCD with $N_f=2$ sea quark flavours, employing the
non-perturbatively improved Wilson-Sheikholeslami-Wohlert action
at a pseudoscalar mass $m_{\pi}= 157(6)$~MeV.
The result is converted
non-perturbatively to the RI'-MOM scheme and then
perturbatively to the $\overline{\mathrm{MS}}$ scheme at a scale $\mu=2$~GeV.
As the quark mass is reduced we find the lattice prediction to
approach the value extracted from
experiments.
\end{abstract}
\maketitle

Almost all visible matter is composed of protons and neutrons.
Analyses of
the scattering of cosmic ray particles off nuclei or of
results from
fixed target and colliding hadron beam experiments
require a quantitative
understanding of the partonic structure of nucleons.
The theoretical framework for this is known~\cite{Georgi:1951sr,Gross:1974cs}
since
the inception of quantum chromodynamics (QCD);
see also~\cite{Christ:1972ms,Callan:1973pu}.

Of particular importance for the experimental
programmes at the Large Hadron Collider are unpolarized
parton distribution functions (PDFs). These, in the
lightcone frame, parameterize
the likelihood of a parton to carry the Bjorken momentum fraction
$x$ at a renormalization
scale $\mu$. While these PDFs have been mapped out very well 
from fits to experimental data,
ideally one would wish to evaluate them directly
from the underlying fundamental theory, QCD.

The present method of choice is lattice QCD,
where in principle all approximations can be removed and systematic
uncertainties controlled by taking the limits of infinite volume,
of vanishing lattice spacing ($a\rightarrow 0$) and of physical
quark masses.
However, in this Monte Carlo
simulation approach to QCD,
the statistical errors and the reliability of the extrapolation
to the physical point are limited by
the power of available
computers and the efficiency of numerical algorithms.
Moreover,
only Mellin moments of the PDFs can be accessed. Thus,
present-day lattice simulation
cannot compete in terms of precision
with determinations of isovector
unpolarized PDFs
from fits to experimental photon-nucleon scattering data that have been
collected over decades of dedicated effort. For
a summary of the present status of PDF parametrizations,
see~\cite{Alekhin:2011sk}.

The possibility to predict averages over the momentum fraction,
however,
is complementary to experimental measurements that can only
cover a limited range of $x$ values.
In particular, the strangeness and gluonic PDFs are determined rather
indirectly from experiment, with large uncertainties~\cite{Aad:2012sb}.
Consequently, lattice QCD input is already on the verge of
becoming essential to constrain these and other less well known quantities,
e.g., the pion nucleon $\sigma$
term~\cite{sternbeck,Bali:2011ks,Durr:2011mp,Dinter:2012tt,Freeman:2012ry},
the strangeness fraction of the mass of the proton
$f_{T_s}$~\cite{Bali:2011ks,Durr:2011mp,Dinter:2012tt,Freeman:2012ry,Gong:2012nw,Babich:2010at},
the strange quark contribution to the spin of the
proton
$\Delta s +\Delta\bar{s}$~\cite{Babich:2010at,QCDSF:2011aa}
or individual (valence and sea) quark
contributions to the proton's momentum $\langle x\rangle_u$,
$\langle x\rangle_d$ and $\langle x\rangle_s$~\cite{Liu:2012nz}.
Naturally, for lattice predictions of such quantities to be trusted,
lattice QCD needs to demonstrate
its ability to reproduce known observables within
similar or smaller errors.

We focus on the second moment of the isovector
PDF of the proton $\langle x\rangle_{u-d}$, i.e., on
the difference between the average
momentum fractions carried by the up and the down quarks.
In the isospin symmetric limit, disconnected quark line contributions
cancel in this quantity, which makes it particularly
accessible to lattice simulations. Nevertheless, at present $\langle x\rangle_{u-d}$
seems to be the one benchmark lattice observable that
is least reliably determined.

\begin{figure}[t]
\centerline{
\includegraphics[width=.48\textwidth,clip=]{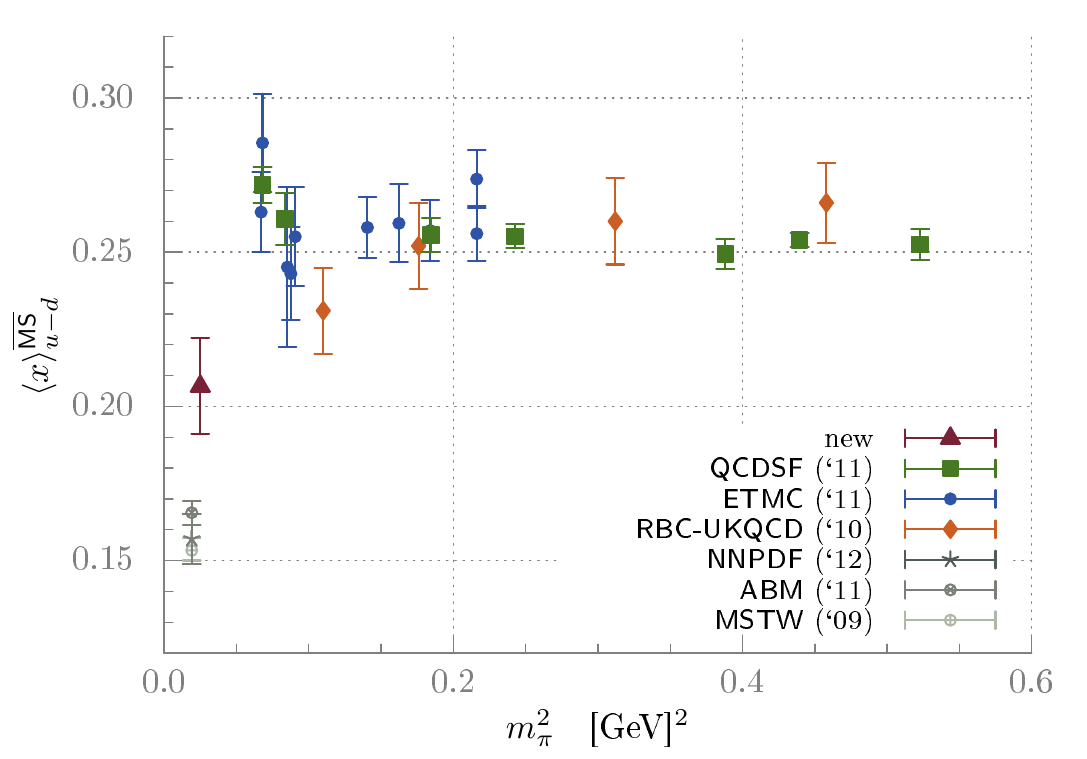}}
\caption{Pion mass dependence of $\langle x\rangle_{u-d}^{\overline{\mathrm{MS}}}$ at $\mu=2$~GeV
from $N_f=2$ (new, QCDSF~\cite{Sternbeck:2012rw,Pleiter:2011gw}, ETMC~\cite{Alexandrou:2011nr}) and
$N_f=2+1$ (RBC-UKQCD~\cite{Aoki:2010xg}) lattice QCD simulations,
together with expectations
from PDF parametrizations (NNPDF~\cite{Ball:2011uy}, ABM~\cite{Alekhin:2012ig}, MSTW~\cite{Martin:2009bu}).
\label{fig:xud}}
\end{figure}

We summarize the present status in Fig.~\ref{fig:xud} where we
plot $\langle x\rangle_{u-d}$ in the $\overline{\mathrm{MS}}$ scheme,
at a scale $\mu=2$~GeV, as a function of the squared pion mass,
obtained by different lattice collaborations. Simulations at
physically light pion masses are particularly expensive and
have only recently become possible for calculating
relatively simple observables, e.g.,
the light hadron spectrum~\cite{Fodor:2012gf}.

The left-most points
(at the physical pion mass $m_{\pi}\approx 138\,\mathrm{MeV}$)
are obtained from recent PDF
parametrizations~\cite{Ball:2011uy,Alekhin:2012ig,Martin:2009bu},
evolved to the renormalization scale $\mu=2$~GeV.
The data point next to it (triangle)
is the main result of this report.
The QCDSF Collaboration points~\cite{Sternbeck:2012rw,Pleiter:2011gw}
stem from simulations
employing the same non-perturbatively
improved~\cite{Jansen:1998mx}
Sheikholeslami-Wohlert $N_f=2$ quark action at lattice scales
ranging from $a^{-1}\approx 2.60$~GeV to $a^{-1}\approx 3.26$~GeV.
The ETMC points~\cite{Alexandrou:2011nr}
were obtained using the $N_f=2$ twisted mass action at maximal twist
at $a^{-1}\approx 2.22$~GeV and at $a^{-1}\approx 3.52$~GeV while
the RBC-UKQCD Collaboration~\cite{Aoki:2010xg} simulated $N_f=2+1$ flavours using the
(approximately) chiral domain-wall action at $a^{-1}\approx 1.7$~GeV.
We have omitted data where the renormalization is not
exactly known~\cite{Bratt:2010jn} from the figure;
see~\cite{Alexandrou:2010cm,Renner:2011rc} for recent reviews.
All the previous data shown in the figure
are compatible with a constant $\sim 0.25$, far above
the expected value $\sim 0.16$ and
no dependence on the different (quark and gluon) lattice
actions, all accurate to $\mathcal{O}(a)$,
on the lattice spacing, on the lattice volume, on including
a strange sea quark or not
or on the pion mass can be resolved within statistical errors.

Covariant baryon chiral perturbation theory suggests
$\langle x\rangle_{u-d}$ to decrease as the physical
point is approached, see, e.g.~\cite{Dorati:2007bk}.
However, the on-set of this behaviour is not at all visible
for $m_{\pi}>250$~MeV, with the possible (but not significant)
exception of the RBC-UKQCD point with the smallest mass value.
Different sources of systematics
in lattice extractions of various nucleon matrix
elements have been discussed in the past.
In particular, contaminations of the ground state
signal, due to excited state contributions,
have recently gained prominence in the
literature~\cite{Green:2011fg,Bulava:2011yz,Capitani:2010sg,Dinter:2011sg}:
reducing these pollutions results in
smaller $\langle x\rangle_{u-d}$ predictions.
This effect, though noticeable, however seems to be too small to
dominantly contribute
to the deviations by factors of about 1.5
of present lattice results from values obtained
from PDF parametrizations.

We report on our new result that is obtained simulating
$N_f=2$ non-perturbatively
improved Sheikholeslami-Wohlert fermions on top of the
Wilson gauge action at $\beta=5.29$
and $\kappa=0.13640$,
corresponding to $m_{\pi}=157(6)\,\mathrm{MeV}$,
on a volume of
$48^3\times 64$ lattice points.
Setting the scale from the (chirally extrapolated)
nucleon mass
we obtain the inverse lattice spacing~\cite{sternbeck}
$a^{-1}=2.76(5)(6)\,\mathrm{GeV}$,
where the errors are statistical and from the chiral extrapolation,
respectively.
Our spatial linear lattice dimension $L\approx 3.43$~fm
is rather small
in units of the pion mass: $m_{\pi}L=2.74(3)$.
However, in previous simulations at
$m_{\pi}\approx 290$~MeV,
differences between results from three volumes
of linear sizes
$m_{\pi}L\approx 2.5, 3.4$ and 4.2 could not be
resolved~\cite{Sternbeck:2012rw},
within statistical errors smaller than the present one.
Hence, we expect finite volume effects to be much smaller
than our statistical uncertainty.

The second Mellin moment of the isovector quark distribution
in the proton is given by
\begin{equation}
\langle x\rangle_{u-d}=\int_0^1\!dx\,x\left[u(x)+\bar{u}(x)-d(x)-\bar{d}(x)\right]\,,
\end{equation}
where $u(x)$, $d(x)$ and $\bar{u}(x)$, $\bar{d}(x)$ denote the quark
and antiquark PDFs, respectively.
It is computed by 
creating a proton at a Euclidean time $t_0=0$, destroying it at a time
$t_f>0$ and inserting
the current
\begin{equation}
\bar{u}\left(\gamma_4\!\stackrel{\leftrightarrow}{D}_4 -\, \frac13\boldsymbol{\gamma}
\cdot\!\stackrel{\leftrightarrow}{\mathbf{D}}\right)u - 
\bar{d}\left(\gamma_4\!\stackrel{\leftrightarrow}{D}_4 -\,\frac13\boldsymbol{\gamma}
\cdot\!\stackrel{\leftrightarrow}{\mathbf{D}}\right)d\,,
\end{equation}
projected to zero spatial momentum,
at an intermediate time $t < t_f$. For details see, e.g.,
the review~\cite{Hagler:2009ni}.
We use mass-degenerate,
electrically neutral $u$ and $d$ quarks and therefore all
disconnected quark line loops cancel.
The resulting three-point function
is normalized by dividing out the two-point function
of the propagating proton. Keeping $t_f$ fixed and sufficiently large,
for $t_f\gg t\gg 0$
one obtains a plateau in $t$ from which the lattice matrix
element can be extracted. 
This
is then translated into the $\overline{\mathrm{MS}}$ scheme
at a scale $\mu=2\,\mathrm{GeV}$, using the renormalization factor
determined in~\cite{Gockeler:2010yr}, which contains non-perturbative
renormalization into the intermediate RI'-MOM scheme and
a subsequent perturbative conversion at three-loop accuracy
into the $\overline{\mathrm{MS}}$ scheme.

\begin{figure}[t]
\centerline{
\includegraphics[width=.48\textwidth,clip=]{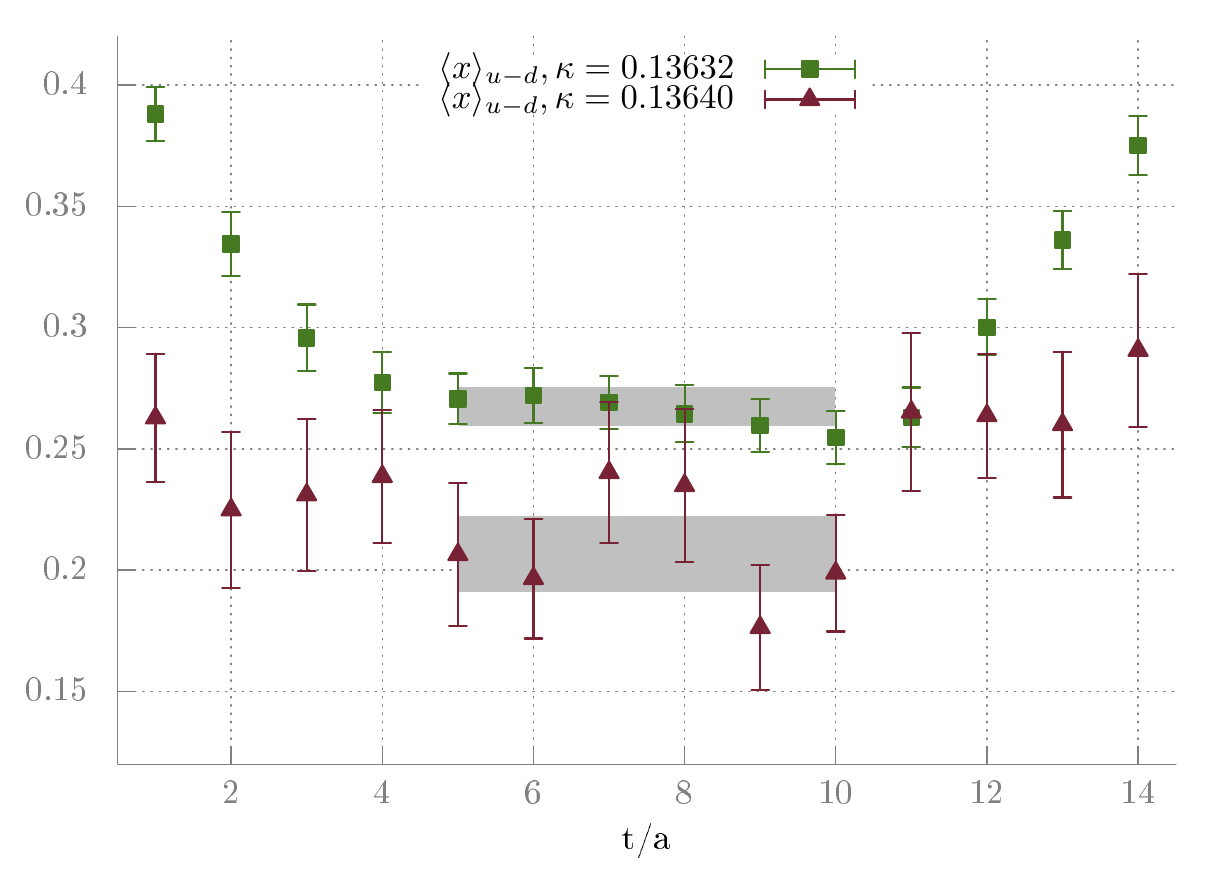}}
\caption{Plateaus and fitted values (grey bands) of ratios of
renormalized
three-point over two-point functions at $\beta=5.29$ ($a^{-1}\approx 2.76$~GeV)
for $\kappa=0.13632$ ($m_{\pi}\approx 290$~MeV) and
$\kappa=0.13640$ ($m_{\pi}\approx 157$~MeV).
\label{fig:plateau}}
\end{figure}
In Fig.~\ref{fig:plateau} we show the resulting plateau and fit
to the renormalized $\langle x\rangle_{u-d}^{\overline{\mathrm{MS}}}$
where we use
$t_f=15a\approx 1.07$~fm. We also compare this to our
previous result obtained for the same
lattice spacing at $m_{\pi}\approx 290$~MeV
($\kappa=0.13632$), on a $40^3\times 64$ lattice~\cite{Sternbeck:2012rw}. 
The on-set and the quality of the plateau depend on the overlap of
the interpolating field used to create and to destroy the proton
with the physical ground state.
Details of our quark and gauge field smearing
can be found in~\cite{Bali:2011ks}. Here we set the number of
Wuppertal smearing iterations to 400.
The larger pion mass data~\cite{Sternbeck:2012rw} were generated with 
sources and sinks of a smaller (Jacobi) smearing radius
and exhibit more curvature as a function of $t/a$,
indicating that the ground 
state overlap is inferior in this case.
This is also obvious from a comparison of the respective
two-point functions. Improving the smearing,
varying $t_f$ in addition to $t$ and
employing more sophisticated fit functions may 
somewhat reduce previous
values that were obtained at the larger
pion masses. Initial tests however indicate that
the difference in quality of the interpolators used to
create the proton can only explain a fraction of
the observed reduction
of $\langle x\rangle_{u-d}$ as the quark mass is decreased.

\begin{figure}[t]
\centerline{
\includegraphics[width=.48\textwidth,clip=]{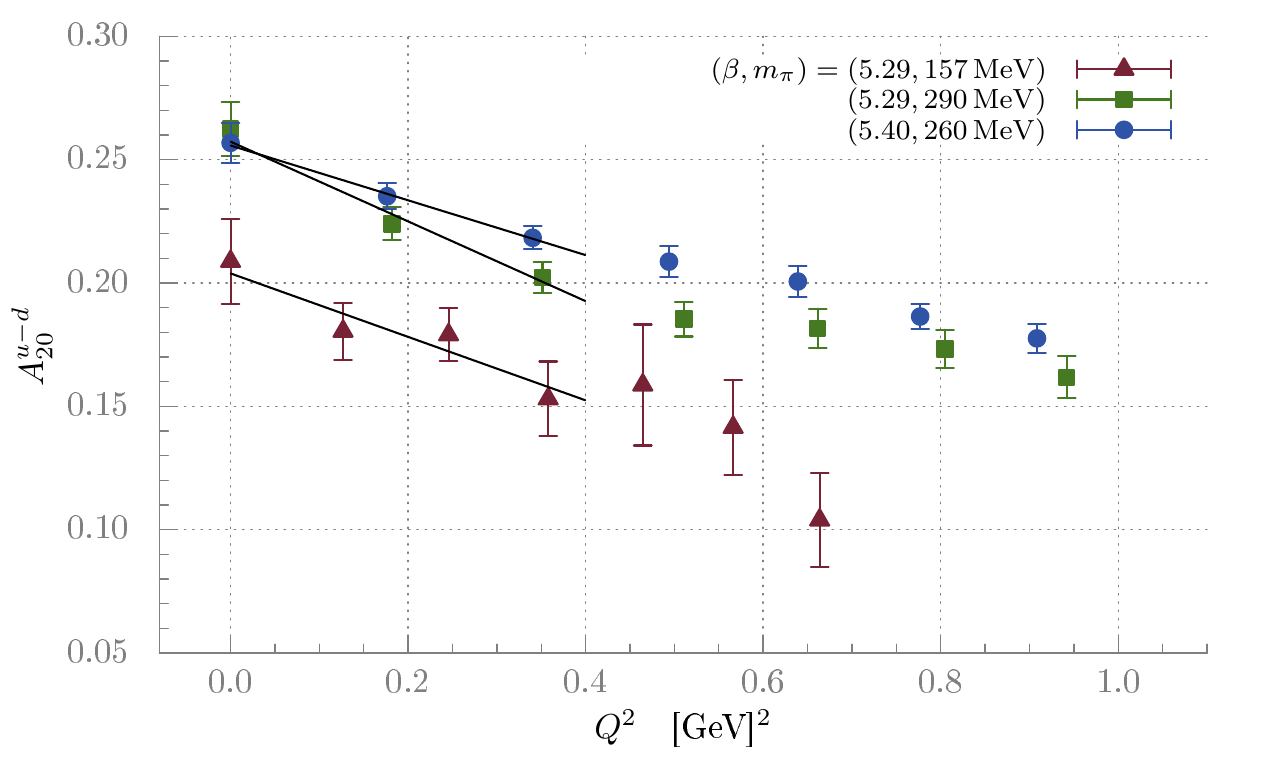}}
\caption{Extrapolations of the generalized form factor $A_{20}^{u-d}(Q^2)$
to $Q=0$ for the three smallest mass points. The left-most
points ($Q=0$) are determined directly from the forward matrix element.
\label{fig:textra}}
\end{figure}

In Fig.~\ref{fig:textra} we display results for the
generalized form factor $A_{20}^{u-d}$ as a function
of the squared
momentum transfer $Q^2$ for the present mass point and
for the previous two smallest QCDSF pion masses~\cite{Sternbeck:2012rw}.
($\beta=5.4$ corresponds to the somewhat
finer lattice scale $a^{-1}\approx 3.26$~GeV.)
To leading order in chiral perturbation theory
these are expected to extrapolate
linearly in $Q^2$ to $\langle x\rangle_{u-d}=A_{20}^{u-d}(0)$,
see, e.g.,~\cite{Dorati:2007bk,Hagler:2009ni}.
The results agree with those obtained
from the forward matrix elements
that are also displayed in Fig.~\ref{fig:xud}
(left-most points).

This report is based on a (computationally expensive) analysis of the first
1000 configurations of our recent $N_f=2$ simulations
at a pion mass $m_{\pi}\approx 157$~MeV and at an inverse lattice spacing
$a^{-1}\approx 2.76$~GeV. Altogether, about 3500 configurations 
have been generated at these
parameter values on the QPACE supercomputers~\cite{Baier:2009yq}.
Hence, we expect the present error to reduce by a factor close to two,
once the target statistics are reached.
We find $\langle x\rangle_{u-d}$ to
drop substantially, an effect that was not visible at $m_{\pi}>250$~MeV.
Part of this may be attributed to some excited state
pollution of the previous data points. A re-analysis of these is also
in progress.

Our value at $m_{\pi}\approx 157$~MeV,
\begin{equation}
\langle x\rangle_{u-d}^{\overline{\mathrm{MS}}}(2\,\mathrm{GeV})=0.207(16)\,,
\end{equation}
is still by 2.5 standard deviations larger than the
central values of
PDF parametrizations~\cite{Alekhin:2011sk,Ball:2011uy,Alekhin:2012ig,Martin:2009bu}. Fig.~\ref{fig:xud}, however, suggests $\langle x\rangle_{u-d}$ to 
decrease steeply
as the physical pion mass is approached.
Moreover, our error, that incorporates the insignificant renormalization
uncertainty~\cite{Gockeler:2010yr}, is purely statistical otherwise.
It does not reflect any systematics of the missing
continuum limit and infinite volume extrapolations.
We remark that the comparison between lattice QCD results and
fits to experimental data for
$\langle x\rangle_{u-d}$ is less clean than for 
the individual moments $\langle x\rangle_u$ and $\langle x\rangle_d$.
These, however, would require disconnected quark line diagrams to
be computed on the lattice.
Experimentally, the neutron is usually
bound in a deuteron or another nucleus.
The EMC effect~\cite{Aubert:1983rq}
will affect PDFs at large $x$~\cite{Accardi:2011fa} and change
their moments, relative to those of a free neutron.
On the lattice side, isospin symmetry was
assumed. However, a strong slope as a function of the pion
mass may also indicate significant corrections to this
approximation. We plan to investigate this further.

We conclude that
lattice simulations are likely to
reproduce known moments of unpolarized PDFs in the near future.
This prediction is based on our observation of a steep decrease
of the value of $\langle x\rangle_{u-d}$
towards small pion masses.
Clearly, high statistics simulations at pion masses smaller than
200~MeV and with good control over excited state contributions
are required.

We thank Y.\ Nakamura and J.\ Zanotti
for their support.
This work was
funded by the DFG Sonderforschungsbereich/Transregio 55 in part
and is supported by the EU Initial Training Network STRONGnet 238353. 
S.\ Collins
acknowledges support from the Claussen-Simon-Foundation (Stifterverband
f\"ur die Deutsche Wissenschaft) and A.\ Sternbeck from the EU (IRG 256594).
Computations were performed on the
SFB/TR55 QPACE computers~\cite{Baier:2009yq},
the BlueGene/P (JuGene) of the J\"ulich Supercomputing Centre,
the SuperMIG/MUC machine of the Leibniz-Rechenzentrum Munich
and Regensburg's iDataCool cluster.
The {\sc Chroma} software suite~\cite{Edwards:2004sx} was used and
gauge configurations were generated with the {\sc BQCD}
code~\cite{Nakamura:2010qh}.

\end{document}